\documentclass[
  aps,
  prl,
  twocolumn,
  superscriptaddress,
  nofootinbib,
  longbibliography,
  fleqn
]{revtex4-2}

\usepackage{amsmath}
\usepackage{amssymb}
\usepackage{mathtools}
\usepackage{mathrsfs}
\usepackage{dsfont}
\usepackage{upgreek}
\usepackage{xspace}
\usepackage{microtype}
\usepackage{xcolor}
\usepackage{acro}
\usepackage{tikz}
\usetikzlibrary{arrows.meta,bending,positioning,decorations.pathreplacing,calligraphy}
\usepackage[hidelinks]{hyperref}
\usepackage[capitalise,noabbrev]{cleveref}

\newcommand{\dd}{\mathop{}\!\mathrm{d}}
\newcommand{\ii}{\mathop{}\!\mathrm{i}\!\mathop{}}
\newcommand{\ee}{\mathrm{e}}
\newcommand{\piup}{\uppi}
\newcommand{\Group}[2]{\mathrm{#1}\!\bigl(#2\bigr)}

\newcommand{\rep}[1]{\boldsymbol{#1}}
\newcommand{\CP}{\ensuremath{\mathcal{CP}}\xspace}
\newcommand{\defeq}{\mathrel{\mathop:}=}
\newcommand{\ASUGRA}{A_\mathrm{SG}}
\DeclareMathOperator{\re}{Re}
\DeclareMathOperator{\im}{Im}

\DeclareMathOperator{\diag}{diag}
\DeclareMathOperator{\Div}{div}
\makeatletter
\g@addto@macro\bfseries{\boldmath}
\makeatother
\definecolor{c1}{RGB}{0,119,187}
\definecolor{c2}{RGB}{51,187,238}
\definecolor{c3}{RGB}{0,153,136}
\definecolor{c4}{RGB}{238,119,51}
\definecolor{c5}{RGB}{204,51,17}
\definecolor{c6}{RGB}{238,51,119}
\definecolor{c0}{RGB}{187,187,187}

\DeclareAcronym{CG}{
  short = CG,
  long = Clebsch--Gordan
}
\DeclareAcronym{CKM}{
  short = CKM,
  long = Cabibbo--Kobayashi--Maskawa
}
\DeclareAcronym{EFT}{
  short = EFT,
  long = effective field theory
}
\DeclareAcronym{GS}{
  short = GS,
  long = Green--Schwarz
}
\DeclareAcronym{IR}{
  short = IR,
  long = infrared
}
\DeclareAcronym{MSSM}{
  short = MSSM,
  long = minimal supersymmetric standard model
}
\DeclareAcronym{QCD}{
  short = QCD,
  long = quantum chromodynamics
}
\DeclareAcronym{SM}{
  short = SM,
  long = standard model
}
\DeclareAcronym{SUGRA}{
  short = SUGRA,
  long = supergravity
}
\DeclareAcronym{SUSY}{
  short = SUSY,
  long = supersymmetry
}
\DeclareAcronym{UV}{
  short = UV,
  long = ultraviolet
}
\DeclareAcronym{VEV}{
  short = VEV,
  long = vacuum expectation value
}
\DeclareAcronym{VVMF}{
  short = VVMF,
  long = vector-valued modular form
}

\begin{document}

\title{Strong \texorpdfstring{\CP}{CP} and Quark Mass Hierarchies from Modular Invariance}

\author{Xiang-Gan Liu}
\email{xianggal@uci.edu}
\affiliation{Department of Physics and Astronomy, University of California, Irvine, California 92697-4575, USA}
\affiliation{Instituto de F\'isica, Universidad Nacional Aut\'onoma de M\'exico, Ciudad de M\'exico C.P. 04510, M\'exico}

\author{Michael Ratz}
\email{mratz@uci.edu}
\affiliation{Department of Physics and Astronomy, University of California, Irvine, California 92697-4575, USA}

\author{Alexander Stewart}
\email{ajstewa1@uci.edu}
\affiliation{Department of Physics and Astronomy, University of California, Irvine, California 92697-4575, USA}

\begin{abstract}
The strong $\CP$ problem and quark mass hierarchies can probe the same ultraviolet structure.
We show this by extending modular strong-$\CP$ solutions to non-Abelian finite modular symmetries.
This reveals a previously overlooked modular-anomaly condition from the finite representations. Regularity removes the dependence of the QCD angle on the $\CP$-breaking modulus, yielding $\bar\theta=0$. 
For positive modular weight, it also forces the quark Yukawa determinant to vanish at the cusp. This leads to pronounced mass hierarchies among the quarks.
We further stress that modular symmetries act as R-symmetries. 
Under moderate additional assumptions, this renders the QCD angle independent of the dilaton. An explicit model illustrates the mechanism.
\end{abstract}

\maketitle

\section{Introduction}\label{sec:introduction}

The strong $\CP$ problem and the flavor puzzle are two persistent structural issues of the \ac{SM} of particle physics. 
The physical \ac{QCD} angle obeys $\lvert\bar\theta\rvert\lesssim10^{-10}$, although no symmetry explains its smallness.
At the same time, the fermion masses span many orders of magnitude and the quark sector exhibits small mixing angles but an unsuppressed weak-\CP phase. 
Conventional solutions often address the strong $\CP$ problem independently of flavor: axion models relax $\bar\theta$ dynamically~\cite{Peccei:1977hh,Weinberg:1977ma,Wilczek:1977pj}, whereas Nelson--Barr models impose exact $\CP$ and a protected structure of the colored mass matrix~\cite{Nelson:1983zb,Barr:1984qx}.

Modular flavor symmetries have been utilized to address the flavor puzzle. 
In this scheme,
Yukawa couplings are \acp{VVMF} of a complex modulus $\tau$, whose transformation properties constrain masses and mixings~\cite{Feruglio:2017spp,Liu:2019khw,Feruglio:2019ybq,Liu:2021gwa,Almumin:2022rml,Kobayashi:2023zzc,Ding:2023htn,Ding:2024ozt}. 
This scheme is motivated by string theory, and can be highly predictive. In particular, it allows us to explain fermion mass hierarchies from the properties of the \acp{VVMF}~\cite{Feruglio:2021dte,Novichkov:2021evw,Chen:2025tby}. 
The same modulus can break $\CP$ spontaneously and yield a nonzero weak $\CP$ phase~\cite{Dent:2001cc,Novichkov:2019sqv,Baur:2019kwi,Ding:2021iqp}.

The observation that target-space modular invariance bears on the strong $\CP$ problem dates back to early four-dimensional string analyses~\cite{Ibanez:1991qh,Kobayashi:2020oji}, and has recently been developed into explicit solutions~\cite{Feruglio:2023uof,Feruglio:2024ytl,Petcov:2024vph,Penedo:2024gtb,Feruglio:2025ajb}. 
We propose a nontrivial extension of the basic mechanism in three directions. 
First, we retain general finite-image representations $\rho_I(\gamma)$ of $\Group{SL}{2,\mathds{Z}}$. 
This allows non-Abelian finite modular multiplets and low-weight \acp{VVMF}, which are natural for twisted string states and essential for the parameter economy expected of modular flavor models. 

Second, there are important contributions to the anomaly coefficients coming from $\rho_I(\gamma)$. 
A matter transformation contains both the nonconstant factor $j_\gamma(\tau)^{-k_I}$ and the finite matrix $\rho_I(\gamma)$, with both contributing to the Jacobian of the path integral measure transformation. 
We find that this leads to an additional constraint which was not part of the classic analyses of target-space modular anomalies~\cite{Derendinger:1991hq,LopesCardoso:1992yd,Ibanez:1992hc}.
Since modular symmetry is a gauge symmetry, this is a genuine quantum-consistency condition; its general implications will be discussed in detail in~\cite{CLLMRS:2026anomaly}. 
As we shall see, this leads to important modifications of the solution of the strong $\CP$ problem.

Third, we will show that in $\mathcal{N}=1$ \ac{SUGRA} $\bar\theta$ is the phase of a holomorphic quantity, $\ASUGRA$, that we are going to define. 
The requirement that $\ASUGRA$ be regular leads to mass hierarchies. 
This is a major qualitative difference from previous proposals \cite{Feruglio:2023uof,Feruglio:2024ytl}, where the hierarchies do not come from \acp{VVMF} but rather from the K\"ahler potential.

We will illustrate our mechanism with a modular $\Delta(384)$ quark model. 

\section{Framework}\label{sec:framework}

We consider a four-dimensional $\mathcal{N}=1$ \ac{SUGRA} \ac{EFT} with a modulus $\tau$, the chiral dilaton $S$, matter multiplets $\Phi^{(I)}$, and Wilsonian gauge kinetic functions $f_a$ of gauge factors $G_a$. In minimal superspace, the Lagrange density is given by
\begin{align}\label{eq:sugra-action}
\mathscr{L}={}&{}-3\int\dd^4\theta\,E\,\ee^{-\mathscr{K}/3}\\
&{}+\left[\int\dd^2\theta\,2\mathcal{E}
\left(\mathscr{W}+\frac{1}{4} f_a\,W^{a\alpha}W^a_{\alpha}\right)+\mathrm{h.c.}\right]\;,\notag
\end{align}
with $g_a^{-2}=\re f_a$ and $\theta_a=-8\piup^2\im f_a$. 
Here, $\mathscr{K}$, $\mathscr{W}$, $E$ and $2\mathcal E$ denote the K\"ahler, superpotential, invariant full- and chiral-superspace densities, respectively. $\theta$ stands for the superspace coordinates, and $W^{a\alpha}W^a_{\alpha}$ for the supersymmetric field strength.
At tree level,
\begin{subequations}\label{eq:tree-data}
\begin{align}
\label{eq:Kahler}
\mathscr{K}={}&-\ln(S+\bar S)-\ln(-\ii\tau+\ii\bar\tau) \notag\\
&{}+\sum_I\frac{\Phi^{(I)\,\dagger}\Phi^{(I)}}{(2\im\tau)^{k_I}}+\cdots\;,\\
\label{eq:superpotential}
\mathscr{W}={}&\sum_nY_{I_1\cdots I_n}(\tau)\,\Phi^{(I_1)}\cdots\Phi^{(I_n)}+\mathscr{W}_\mathrm{hid}\;,\\
f_a^{(0)}={}&\kappa_a\,S\;.
\end{align}
\end{subequations}
Here, $Y_{I_1\cdots I_n}(\tau)$ denotes the holomorphic Yukawa coupling depending on $\tau$, and $\mathscr{W}_\mathrm{hid}$ stands for additional superpotential terms, including those stemming from a so-called hidden sector.
The diagonal matter metric in \Cref{eq:Kahler} is chosen only to fix our
modular-weight convention; the strong-$\CP$ argument below does not
rely on this special minimal form. At one-loop, the \ac{GS} completion makes $S+\bar S-\delta_\mathrm{GS}(8\piup^2)^{-1}\ln(2\im\tau)$ modular invariant.

\subsection{Modular Invariance}\label{sec:modular-invariance}

For $\gamma=\bigl(\begin{smallmatrix}a&b\\c&d\end{smallmatrix}\bigr)\in\Group{SL}{2,\mathds{Z}}$, let $j_\gamma(\tau)\defeq c\,\tau+d$ and $\Sigma_I(\gamma,\tau)\defeq j_\gamma(\tau)^{-k_I}\,\rho_I(\gamma)$ with $\rho_I(\gamma)$ denoting the representation matrix under the so-called finite modular group. 
For one-dimensional $\rho_I$, we write $\chi_I\coloneq\rho_I$.
We adopt the explicit \ac{GS} frame described in \Cref{app:GS-frame} of the Supplemental Material, and the modular transformations of the relevant superfields are given by
\begin{subequations}\label{eq:modular-transformations}
\begin{align}
\tau&\xmapsto{~\gamma~}\gamma\,\tau=\frac{a\,\tau+b}{c\,\tau+d}\;,\label{eq:modular-transformations_tau}\\
S&\xmapsto{~\gamma~} S-\frac{\delta_\mathrm{GS}}{8\piup^2}\ln j_\gamma(\tau)\;,\label{eq:S-GS}\\
\Phi^{(I)}&\xmapsto{~\gamma~} \Sigma_I(\gamma\,,\tau)\,\Phi^{(I)}\;.\label{eq:matter-transform}
\end{align}
\end{subequations}
The K\"ahler potential of $\tau$ is not invariant but rather shifts by $F_\gamma+\bar F_\gamma$, with $F_\gamma\defeq k_{\mathscr{W}}\ln j_\gamma$ and $k_{\mathscr{W}}=1$~\cite{Dixon:1989fj,Ferrara:1989qb}. This shift is compensated by the corresponding transformation of the superpotential,
\begin{equation}\label{eq:W-transform}
\mathscr{W}\xmapsto{~\gamma~} j_\gamma^{-k_{\mathscr{W}}}\,\mathscr{W}\;.
\end{equation}
This implies that modular symmetries are necessarily R-symmetries. 
We stress that this conclusion is unavoidable in supersymmetric modular invariant theories since the only way to make the scheme consistent with gravity is \ac{SUGRA}, where \eqref{eq:W-transform} is dictated by modular invariance. 
As we shall see, this fact has important implications for the solution of the strong $\CP$ problem.
An important additional implication is that, since the modulus is inert under $\gamma=\mathsf{S}^2=-\mathds{1}_2$, every modular invariant theory has a linearly realized symmetry that maps  
\begin{equation}\label{eq:Z4R}
  \mathscr{W}\xmapsto{~\mathsf{S}^2~}-\mathscr{W}\;.
\end{equation}
The superspace coordinates pick up $\pm\ii$, so this is in fact a linearly realized order-4 R-symmetry, $\mathds{Z}_4^R$.

We choose the full superpotential to transform trivially under the finite modular group as shown in \Cref{eq:W-transform}. 
The holomorphic coefficient $Y_{I_1\cdots I_n}(\tau)$ of a monomial $\Phi^{(I_1)}\cdots\Phi^{(I_n)}$ in the superpotential \eqref{eq:superpotential} must then obey
\begin{equation}\label{eq:Y-VVMF-transform}
Y_{I_1\cdots I_n}(\gamma\,\tau)
=
j_\gamma(\tau)^{k_Y}\rho_Y(\gamma)\,Y_{I_1\cdots I_n}(\tau)
\end{equation}
with $k_Y=\sum k_{I_i}-k_{\mathscr{W}}$ and $\rho_Y\otimes\rho_{I_1}\otimes\cdots\otimes\rho_{I_n}\supset\rep 1$. Thus, $\Group{SL}{2,\mathds{Z}}$ invariance forces the Yukawa couplings to be highly constrained \acp{VVMF}~\cite{Feruglio:2017spp,Liu:2021gwa}, rather than arbitrary holomorphic functions.

For target-space modular symmetries arising from string compactifications, modular transformations are discrete gauge identifications, so $\tau$ and $\gamma\tau$ describe the same physical vacuum. The inequivalent vacua are therefore parametrized by the fundamental domain $\Group{SL}{2,\mathds{Z}}\backslash\mathds{H}$, including the cusp
\begin{equation}\label{eq:modular-curve}
\mathcal{F}=\Group{SL}{2,\mathds{Z}}\backslash\mathds{H}^*~\text{with}~
\mathds{H}^*=\mathds{H}\cup\mathds{Q}\cup\{\ii\infty\}\;,
\end{equation}
which is also known as the compactified modular curve.

\subsection{Two Layers of the Modular Anomaly}\label{sec:modular-anomaly}

The fermion measure is not invariant under \Cref{eq:matter-transform}. 
For a multiplet $I$ in a gauge representation $\rep{R}^{(I)}$ of $G_a$, let $d_I=\dim\rho_I$ and let $m_I^{(a)}$ count gauge multiplicities.
The transformation determinant is
\begin{equation}\label{eq:det-Sigma}
\det\Sigma_I^{m_I^{(a)}}
=j_\gamma^{-k_I\,d_I\,m_I^{(a)}}\det\rho_I(\gamma)^{m_I^{(a)}}\;.
\end{equation}
With the convention that $\mathcal{A}_a$ is the coefficient required in the compensating variation of $f_a$, the light-spectrum anomaly data are
\begin{align}
\mathcal{A}_a&=k_{\mathscr{W}}
\left[C_a-\sum_I\ell_a(\rep{R}^{(I)})\,d_I\,m_I^{(a)}\right]
\nonumber\\
&\quad{}+\sum_I2\ell_a(\rep{R}^{(I)})\,d_I\,m_I^{(a)}\,k_I\;,
\label{eq:anomaly-coefficient}\\
\chi_a^\rho(\gamma)&=\prod_I
\det\rho_I(\gamma)^{2\ell_a(\rep{R}^{(I)})\,m_I^{(a)}}\;,
\label{eq:finite-character}
\end{align}
where $C_a$ denotes the quadratic Casimir of $G_a$, and $\ell_a$ the Dynkin indices. Since determinants are multiplicative,
$\chi_a^\rho(\gamma)$ defines a one-dimensional determinant
character of the modular group, encoding the finite phase of the
fermion Jacobian.
The Wilsonian coupling must consequently satisfy
\begin{subequations}\label{eq:gauge-transform}
\begin{align}
\Delta f_a& \defeq f_a(S',\gamma\,\tau)-f_a(S,\tau)\notag\\ 
&=\frac{\mathcal{A}_a}{8\piup^2}\ln j_\gamma
-\frac{1}{8\piup^2}\ln\chi_a^\rho(\gamma)\;,\\
\ee^{-8\piup^2 f_a(S',\gamma\,\tau)}
&=j_\gamma^{-\mathcal{A}_a}\,\chi_a^\rho(\gamma)\,
\ee^{-8\piup^2 f_a(S,\tau)}\;.
\label{eq:instanton-transform}
\end{align}
\end{subequations}

The one-loop holomorphic coupling in the \ac{GS} frame is
\begin{equation}\label{eq:threshold-form}
f_a(S,\tau)=\kappa_a\,S+
\frac{\widetilde{B}_a}{8\piup^2}\ln\eta(\tau)^2+f_a^\mathrm{inv}(\tau)\;.
\end{equation}
Here $\widetilde{B}_a$ denotes the coefficient of the holomorphic
one-loop threshold. Using $\eta(\gamma\,\tau)^2=\chi_\eta(\gamma)j_\gamma(\tau)\eta(\tau)^2$ gives two separate \ac{UV} modular anomaly cancellation conditions,
\begin{subequations}\label{eq:two-layer-condition}
\begin{align}
\mathcal{A}_a&=\widetilde{B}_a-\kappa_a\,\delta_\mathrm{GS}\;,\label{eq:continuous-layer}\\
1&=\chi_a^\rho(\gamma)\,\chi_\eta(\gamma)^{\widetilde{B}_a}\;,\quad
\forall\gamma\in\Group{SL}{2,\mathds{Z}}\;.
\label{eq:character-layer}
\end{align}
\end{subequations}
The first equation matches the nonconstant automorphy factor and fixes only the combination $\widetilde{B}_a-\kappa_a\,\delta_\mathrm{GS}$. 
The second matches the determinant character to the threshold multiplier. 
This character condition was absent from early analyses~\cite{Derendinger:1991hq,LopesCardoso:1992yd,Ibanez:1992hc}.\footnote{Since every one-dimensional character of $\Group{SL}{2,\mathds{Z}}$ is a power of $\chi_\eta$, we may write $\chi_a^\rho=\chi_\eta^{n_a}$ with $n_a\in\mathds{Z}_{12}$. 
\Cref{eq:character-layer} then fixes the congruence class
$\widetilde{B}_a\equiv -n_a\pmod{12}$. In particular, $\chi_a^\rho=1$ implies $\widetilde{B}_a\in12\mathds{Z}$.} 
This provides an independent consistency condition on bottom-up modular flavor model building~\cite{CLLMRS:2026anomaly}. 
In a string construction it is instead a necessary consistency condition relating the finite modular representations of the light spectrum to the threshold data of the \ac{UV} completion. 
In heterotic orbifolds, the coefficients $\widetilde{B}_a$ are determined by the corresponding string thresholds, in particular by the $\mathcal{N}=2$ subsector data~\cite{Kaplunovsky:1995jw}. 

In the minimal cusp-only class, regularity in the interior, together
with the standard decompactification behavior
$\re f_a\supset\mathcal{O}(\im\tau)$, excludes additional
nonconstant modular invariant contributions built from
$j(\tau)$~\cite{Kaplunovsky:1995jw}. 
The gauge kinetic function therefore reduces to
\begin{equation}\label{eq:minimal-gauge-function}
f_a(S,\tau)=\kappa_a\,S
+\frac{\widetilde{B}_a}{8\piup^2}\ln\eta(\tau)^2
+\mathrm{constant}\;.
\end{equation}
Here, we represent the so-called threshold correction by $\ln\eta(\tau)^2$, which has a zero at $\tau=\ii\infty$, for the sake of concreteness. 
Our analysis can be extended to apply for more general threshold terms, which may have a richer pole structure.

\section{Solving the strong \texorpdfstring{\CP}{CP} Problem and generating Mass Hierarchies}\label{sec:strong-cp}

After \ac{SUSY} breaking, the physical QCD angle is
\begin{equation}\label{eq:theta-bar}
\bar\theta=-8\piup^2\im f_3+\arg\det(Y^u\,Y^d)+C_3\,\arg M_3\;,
\end{equation}
where $C_3=3$ and $M_3$ denotes the gluino mass.
We assume that the \ac{SUSY} breaking sector preserves \CP and that the gluino mass satisfies $M_3=r_3\,\overline{\mathscr{W}}_0$ where $r_3\in\mathds{R}$, so that
\begin{equation}
\label{eq:gluino-locking}
\arg M_3=-\arg\mathscr{W}_0\bmod\piup\;.
\end{equation}
Note that this condition can be motivated if the relevant derivatives of $f_3$ are real in a $\CP$ basis.
The detailed derivation is given in \Cref{app:phase-hidden} of the Supplemental Material. 
It is convenient to define 
\begin{equation}\label{eq:A-SUGRA}
\ASUGRA\defeq\ee^{-8\piup^2f_3(S,\tau)}\,
\det(Y^u\,Y^d)(\tau)\,\mathscr{W}_0(S,\tau)^{-C_3}\;,
\end{equation}
which depends on $S$ and $\tau$. The phase of this quantity is given by $\bar\theta$, cf.~\Cref{eq:theta-bar}.
The Yukawa determinant transforms with weight
\begin{equation}\label{eq:det-weight}
\mathcal{K}_Y=\sum_{i=1}^3(2k_{Q_i}+k_{U_i^c}+k_{D_i^c})
+3(k_{H_u}+k_{H_d})-6k_{\mathscr{W}}\;.
\end{equation}
Its finite character is $(\chi_3^\rho)^{-1}(\chi_{H_u}\,\chi_{H_d})^{-3}$. 
Combining this with \Cref{eq:instanton-transform,eq:W-transform} gives
\begin{equation}\label{eq:A-transform}
\ASUGRA\xmapsto{~\gamma~}
j_\gamma^{3(k_{H_u}+k_{H_d})}(\chi_{H_u}\chi_{H_d})^{-3}\,\ASUGRA\;.
\end{equation}
We assume that
\begin{equation}\label{eq:Higgs-condition}
k_{H_u}+k_{H_d}=0 ~\text{ and }~ \chi_{H_u}\,\chi_{H_d}=1\;.
\end{equation}
Then $\ASUGRA$ is modular invariant under the combined modular action on $S$ and $\tau$, \eqref{eq:modular-transformations_tau} and \eqref{eq:S-GS}.
Similarly to related proposals \cite{Feruglio:2024ytl}, we require $\ASUGRA$ to be finite throughout moduli space. This implies that $\ASUGRA$ cannot depend on $\tau$,
and further requiring that it is generally nonzero, we have
\begin{align}\label{eq:divisor-condition}
\MoveEqLeft\Div\left[
\ASUGRA
\right] = 0 = \Div\left[\ee^{-8\piup^2f_3(S,\tau)}\right] \\
&{}+\Div\left[\det(Y^u\,Y^d)(\tau)\right]-C_3\Div\left[\mathscr{W}_0(S,\tau)\right] \;,\notag
\end{align}
where the divisor $\Div\left[F\right]\defeq\sum_{p}\operatorname{ord}_{p}(F)\,[p]$ counts the zeros and poles of the function $F$ with their multiplicities.
Crucially, we assume that $\tau$ is the unique source of $\CP$ violation, implying that any $\CP$-violating term involves $\tau$.
As $\ASUGRA$ is independent of the $\CP$-breaking $\tau$, exact $\CP$ symmetry makes the remaining quantity real. Disregarding the option $\bar\theta=\piup$ (cf.\ \cite{Gaiotto:2017yup}), we thus have
\begin{equation}\label{eq:theta_equals_0}
 \bar\theta=\arg \ASUGRA =0\;,
\end{equation}
thereby solving the strong $\CP$ problem.

In our current concrete example, the visible threshold has only the $\eta$ divisor and $\mathscr{W}_0$ has no explicit $\tau$ divisor in the \ac{GS} frame. 
If $\mathcal{K}_Y$ is the weight in \Cref{eq:det-weight}, divisor matching and the valence formula (see e.g.\ \cite{Chen:2025tby} for the definition) give
\begin{equation}\label{eq:cusp-form-condition}
  \widetilde{B}_3=\mathcal{K}_Y 
  ~\text { and }~
  \det(Y^u\,Y^d)\propto\eta^{2\mathcal{K}_Y}\;.
\end{equation}
Note that $\mathcal{K}_Y>0$ implies that the quark masses could exhibit large hierarchies (cf.\ \cite{Feruglio:2021dte,Novichkov:2021evw,Chen:2025tby} for a detailed discussion), and that they depend on $\tau$, thus generically violating $\CP$. 
In order to see this more concretely, we will study an explicit model in what follows. 
The branch $\mathcal{K}_Y=0$ permits a constant determinant, which corresponds to the original modular strong \CP solution framework~\cite{Feruglio:2023uof}.
Let us also mention that, provided that there are no $\Group{SU}{3}$ charged states beyond those of the MSSM and that \Cref{eq:Higgs-condition} holds, $\kappa_3\,\delta_\mathrm{GS}=\mathcal{K}_Y-\mathcal{A}_3=-3k_{\mathscr{W}}$.
Thus the universal $\delta_\mathrm{GS}$ is tied directly to the universal superpotential weight $k_{\mathscr{W}}$. 
The same $\delta_\mathrm{GS}$ consequently constrains hidden gauge sectors, as illustrated below.

\subsection{An explicit Model}\label{sec:model}

Let us consider the finite modular group $\Delta(384)\subset\Gamma_{16}$, and assign the three quark generations according to
\begin{align}
k_Q&=k_{U^c}=k_{D^c}=\tfrac{3}{2}\;,\quad k_{H_u}=k_{H_d}=0\;, \notag\\
\rho_Q&=\rep{3}_0\;,\quad \rho_{U^c}=\rep{3}_9\;,\quad \rho_{D^c}=\rep{3}_{13}\;.\label{eq:model-assignments}
\end{align}
The finite representations and the available \ac{CG} contractions uniquely fix the quark superpotential up to two real constants $\alpha_u$ and $\alpha_d$,
\begin{equation}\label{eq:model-superpotential}
\mathscr{W}_q=\alpha_u\,\bigl(Y^{(2)}_{\rep{6}_0}\,U^c\,Q\bigr)_{\rep{1}_0}\,H_u+\alpha_d\,\bigl(Y^{(2)}_{\rep{6}_5}\,D^c\,Q\bigr)_{\rep{1}_0}\,H_d\;.
\end{equation}
Here $Y^{(2)}_{\rep 6_0}$ and $Y^{(2)}_{\rep 6_5}$ are the weight-2 sextet \acp{VVMF} constructed in \cite{CentellesChulia:2026bkr}. 
The resulting Yukawa matrices are
\begin{subequations}\label{eq:model-Yukawas}
\begin{align}
\frac{Y^u}{\alpha_u}&=
\begin{pmatrix}
-Y^{(2)}_{\rep 6_0,1} & -Y^{(2)}_{\rep 6_0,6} & \sqrt2Y^{(2)}_{\rep 6_0,5}\\
\sqrt2Y^{(2)}_{\rep 6_0,3} & \sqrt2Y^{(2)}_{\rep 6_0,4} & 0\\
Y^{(2)}_{\rep 6_0,6} & Y^{(2)}_{\rep 6_0,1} & \sqrt2Y^{(2)}_{\rep 6_0,2}
\end{pmatrix}\;,\label{eq:Yu-matrix}\\
\frac{Y^d}{\alpha_d}&=
\begin{pmatrix}
Y^{(2)}_{\rep 6_5,5} & Y^{(2)}_{\rep 6_5,6} & -\sqrt2Y^{(2)}_{\rep 6_5,2}\\
-Y^{(2)}_{\rep 6_5,6} & -Y^{(2)}_{\rep 6_5,5} & \sqrt2Y^{(2)}_{\rep 6_5,3}\\
\sqrt2Y^{(2)}_{\rep 6_5,4} & -\sqrt2Y^{(2)}_{\rep 6_5,1} & 0
\end{pmatrix}\;.\label{eq:Yd-matrix}
\end{align}
\end{subequations}
They obey the identity
\begin{equation}\label{eq:model-determinant}
\det(Y^u\,Y^d)=-256\sqrt2\,\alpha_u^3\,\alpha_d^3\,\eta^{24}(\tau)\;.
\end{equation}
The determinant has weight $12$, trivial finite character, and its only zero in the fundamental domain is the cusp.

The color anomaly data of the light spectrum are
\begin{subequations}\label{eq:model-anomaly-data}
\begin{align}
 \mathcal{A}_3&=15\;,\quad \kappa_3=1\;,\\
 \chi_3^\rho&=(\det\rho_Q)^2\,\det\rho_{U^c}\,\det\rho_{D^c}=1\;.
\end{align}
\end{subequations}
Since $\chi_3^\rho$ is trivial, the condition \eqref{eq:character-layer} implies that $\widetilde{B}_3=0\bmod{12}$. 
The weight-$12$, cusp-order-one determinant in \Cref{eq:model-determinant} selects $\widetilde{B}_3=12$ through \Cref{eq:cusp-form-condition}, and the continuous condition \eqref{eq:continuous-layer} implies that $\delta_\mathrm{GS}=-3$. 
Thus,
\begin{equation}\label{eq:model-f3}
  f_3=S+\frac{12}{8\piup^2}\ln\eta^2+c
\end{equation}
with a constant $c$.

The same universal $\delta_\mathrm{GS}$ constrains a minimal hidden sector. 
A threshold-free pure super-Yang--Mills group at level one has $\mathcal{A}_h=C_h$, $\widetilde{B}_h=0$, and therefore $C_h=-\delta_\mathrm{GS}=3$. 
The arguably simplest unitary realization is $\Group{SU}{3}_h$, although other groups with the same dual Coxeter number are possible. 
The standard gaugino-condensation superpotential is
\begin{equation}\label{eq:model-W0}
 \mathscr{W}_0=A\,\exp\left(-\frac{8\piup^2}{3}S\right)\;,
\end{equation}
which transforms as $j_\gamma^{-1}\,\mathscr{W}_0$ with trivial finite character. 
Here, $A$ is a real constant. Substituting \Cref{eq:model-determinant,eq:model-f3,eq:model-W0} into \Cref{eq:A-SUGRA} gives
\begin{equation}\label{eq:model-Aconstant}
\ASUGRA=-256\sqrt{2}\,\ee^{-8\piup^2 c}\alpha_u^3\,\alpha_d^3\,A^{-3}\;.
\end{equation}
The $\tau$ dependence cancels between the visible threshold and the determinant of the quark mass matrices, while the $S$ dependence cancels between the visible gauge factor and the hidden condensate. 
With all the parameters real in a $\CP$ basis, $\ASUGRA$ is real and hence $\bar\theta=0$, cf.~\Cref{eq:theta_equals_0}. The restricted uniqueness and condensate branches are discussed in \Cref{app:phase-hidden} of the Supplemental Material.

Writing $\varepsilon=\lvert q^{1/16}\rvert=\ee^{-\piup\im\tau/8}$, the singular values behave near the cusp as
\begin{subequations}\label{eq:model-hierarchies}
\begin{align}
m_u:m_c:m_t&=4\sqrt2\,\varepsilon^6:2\varepsilon^2:1\;,\\
m_d:m_s:m_b&=2\sqrt2\,\varepsilon^5:\sqrt2\varepsilon^2:\varepsilon\;.
\end{align}
\end{subequations}
Thus large quark-mass hierarchies arise naturally. 
In addition, the bottom Yukawa coupling is further suppressed by one power of $\varepsilon$ relative to the top, $y_b/y_t\sim(\alpha_d/\alpha_u)\varepsilon$. 
A $\CP$ violating \ac{VEV} of $\tau$, i.e.\ for $\re\tau\neq0,\pm1/2$ and $\lvert\tau\rvert\neq 1$~\cite{Dent:2001cc,Novichkov:2019sqv,Baur:2019kwi,Ding:2021iqp},
\begin{equation}\label{eq:Jarlskog-invariant}
J_{\CP}\propto\im\det[Y^{u\dagger}Y^u,Y^{d\dagger}Y^d]\neq0\;.
\end{equation}
Very close to the cusp, the weak-\CP phase is parametrically suppressed.
As the model has four real parameters, $\re\tau$, $\im\tau$, $\alpha_u$ and $\alpha_d$, it is overconstrained and does not yield a perfect fit to observables at tree level. Moderate modular-covariant K\"ahler corrections can change order-one factors without altering the exact strong $\CP$ cancellation or the leading cusp powers.

\section{Corrections and string Origin}\label{sec:corrections-string}

A nonminimal matter K\"ahler metric does not rotate $\bar\theta$. 
Canonical normalization multiplies the holomorphic colored determinant by a positive real function~\cite{Chen:2019ewa},
\begin{equation}\label{eq:kahler-correction}
\widetilde{\ASUGRA}(S,\tau,\bar\tau)=r_Z(\tau,\bar\tau)\,\ASUGRA(S,\tau)
\quad\text{with } r_Z>0\;.
\end{equation}
It can modify masses and mixing angles, but a moderately conditioned metric does not change the leading cusp powers. 
\ac{SM} contributions to $\bar\theta$ arise only at high loop order and are negligible~\cite{Ellis:1978hq,Khriplovich:1985jr,Romanino:1996cn}. 
On the other hand, additional $F$ terms or an uplift sector with independent complex phases may regenerate the ordinary \ac{SUSY} $\CP$ problem~\cite{Dent:2001cc}. 
The pure condensate fixes the modular transformation of $\mathscr{W}_0$ but does not by itself provide a complete model of moduli stabilization, \ac{SUSY} breaking, and vacuum energy tuning. A detailed discussion is left for future work.

An important comment concerns additional vector-like states which are charged under $\Group{SU}{3}_\mathrm{C}$ and acquire a $\tau$-dependent mass $M(\tau)$. 
String constructions often contain such states, cf.~e.g.~\cite{Lebedev:2006kn}.
This give rise to the Kaplunovsky--Louis threshold~\cite{Kaplunovsky:1994fg,Louis:1996mt}
\begin{equation}\label{eq:heavy-threshold-main}
f_3^\mathrm{IR}=f_3^\mathrm{UV}-\frac{1}{8\piup^2}\ln M(\tau)\;.
\end{equation}
This means that the effect of such states is to redefine $f_3$. 
The solution to the strong $\CP$ problem therefore remains unchanged as long as $M(\tau)$ does not contain additional $\CP$ violating phases, which would be inconsistent with our working assumption that $\tau$ is the only source of $\CP$ violation. 
We provide a more detailed discussion in \Cref{app:UVIR} of the Supplemental Material.

The Higgs condition \eqref{eq:Higgs-condition} also constrains the $\mu$ term. 
Modular covariance of $\mathscr{W}\supset\mu(\tau)\,H_u\,H_d$ gives
\begin{equation}\label{eq:mu-transform}
\mu(\gamma\,\tau)=j_\gamma^{k_{H_u}+k_{H_d}-k_{\mathscr{W}}}\mu(\tau)=j_\gamma^{-1}\,\mu(\tau)\;.
\end{equation}
No negative-weight modular forms exist, so the bare superpotential $\mu$ term vanishes as a modular zero~\cite{Liu:2025lym}. 
This is also a consequence of the modular $\mathds{Z}_4^R$ symmetry \eqref{eq:Z4R}, since the Higgs bilinear $H_u\,H_d$ carries R-charge 0.
An appropriate $\mu$ term can then be generated by the Kim--Nilles \cite{Kim:1983dt} and/or Giudice--Masiero \cite{Giudice:1988yz} mechanisms, i.e.\ once the \ac{VEV} of the superpotential breaks the modular R-symmetry or from \ac{SUSY} breaking.

Exact $\CP$ is natural in string theory, where it can be a discrete gauge transformation inherited from higher-dimensional Lorentz symmetry~\cite{Dine:1992ya}. 
Our stronger assumption is that $\tau$ is the only source of complex phase relevant to the colored and supersymmetry-breaking sectors. Extending the construction to several moduli or larger symplectic modular groups~\cite{Ding:2020zxw} requires regularity of the corresponding complete automorphic invariant and the two anomaly conditions for every relevant transformation.

It is instructive to compare our modular solution with the axion and Nelson--Barr solutions. 
Unlike an approximate global Peccei--Quinn symmetry, target-space modular symmetry is an exact gauge redundancy, so the usual quality problem is replaced by the calculable anomaly and divisor conditions. 
As long as $\operatorname{div}\ASUGRA=0$, no light field is required to relax $\bar\theta$. If a residual divisor instead leaves a pseudoscalar modulus dependence, the corresponding direction can provide an axion-like solution, provided that it remains light and its potential is dominated by QCD. 
Similarly to Nelson--Barr models, exact $\CP$ is broken spontaneously while $\bar\theta$ vanishes, but no specially arranged colored quark mass texture is required.

\section{Conclusion}\label{sec:conclusion}

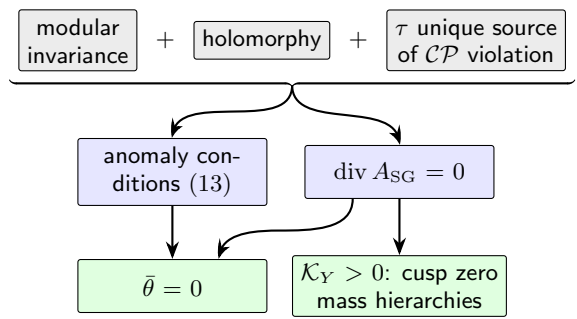
\begin{figure}[th!]
\centering
\begin{tikzpicture}[>={Stealth[bend]},box/.style={rectangle,draw,rounded corners=1pt,align=center,font=\sffamily}]
  \matrix[ampersand replacement=\&,column sep=1ex,row sep=1em,nodes={anchor=center}](mat1){
    \node[box,fill=gray!15]{modular\\ invariance}; \&
    \node{$+$}; \&
    \node[box,fill=gray!15]{holomorphy}; \&
    \node{$+$}; \&
    \node[box,fill=gray!15]{$\tau$ unique source\\ of $\CP$ violation}; \\
  };
  \matrix[below=2em of mat1,ampersand replacement=\&,column sep=1em,row sep=2em,nodes={minimum height=0.7cm,minimum width=2.5cm,anchor=center}](mat2){
    \node[box,fill=blue!10] (anom) {anomaly con-\\ 
    ditions \eqref{eq:two-layer-condition}};
    \&
    \node[box,fill=blue!10] (div) {$\Div \ASUGRA=0$}; \\
    \node[box,fill=green!12] (scp) {$\bar\theta=0$}; 
    \&
    \node[box,fill=green!12
      ] (flavor) {$\mathcal{K}_Y>0$: cusp zero\\mass hierarchies}; \\
  };  
  \draw[thick,decorate,decoration={calligraphic brace}] (mat1.south east) -- coordinate[below=0.5ex](p1) (mat1.south west);
  \path (anom.south east) -- coordinate (p2) (flavor.north west);
  \draw[->,thick] (p1) to[out=-90,in=90] (anom);
  \draw[->,thick] (p1) to[out=-90,in=90] (div);
  \draw[->,thick] (anom)--(scp);
  \draw[->,thick] (div.-150) to[out=-90,in=0] (p2) to[out=180,in=90] (scp.30);
  \draw[->,thick] (div)--(flavor);
\end{tikzpicture}
\caption{Assumptions and implications.}
\label{fig:logic}
\end{figure}

In this Letter, we have revisited the modular solutions of the strong $\CP$ problem.
Like in previous proposals, our main assumptions are modular invariance, that $\tau$ is the unique source of $\CP$ violation, and that the theory is regular throughout moduli space. 
However, our mechanism differs from earlier proposals in three important aspects.
We have stressed that in \ac{SUSY} theories  consistent with gravity, modular symmetries are inevitably R-symmetries. 
In addition, we allowed the quarks to furnish nontrivial finite-image representations $\rho_I(\gamma)$. 
We also find that the duality anomaly constraints have two layers.
The previously overlooked second anomaly condition constrains the
finite modular representations of the light spectrum, providing a new consistency condition for bottom-up modular flavor models. Together with our choice of the Higgs modular weights, these ingredients make $\ASUGRA$, the holomorphic strong-$\CP$ quantity whose phase is $\bar\theta$, modular invariant. Regularity then makes $\ASUGRA$ a $\tau$-independent constant; since $\tau$ is the unique source of $\CP$ violation, exact $\CP$ makes this constant real, yielding $\bar\theta=0$ and thereby solving the strong $\CP$ problem. Furthermore, for $\mathcal{K}_Y>0$, the same regularity condition requires the Yukawa determinant to vanish at the cusp, thereby providing the structural boundary condition for quark-mass hierarchies, as summarized in \Cref{fig:logic}.

We have illustrated the mechanism in a minimal $\Delta(384)$ model, in which $\det(Y^uY^d)\propto\eta^{24}$ implies that $\widetilde{B}_3=12$ and $\delta_\mathrm{GS}=-3$. 
In a minimal threshold-free single-condensate scheme, the same anomaly data select a hidden pure-super-Yang--Mills sector with $C_h=3$, whose condensate even further cancels the remaining dilaton dependence. 
A \ac{VEV} of $\tau$ inside the fundamental domain gives a nontrivial weak $\CP$ violation while ensuring the strong $\CP$ phase is zero. 
Our toy model is overconstrained and does not yield a fully realistic fit to quark observables. 
Nevertheless, it provides a semi-realistic and highly predictive proof of principle for the exact quark Yukawa determinant, anomaly, hidden-sector, and hierarchy relations.

The fact that modular symmetries are R-symmetries implies that the gravitino mass $m_{3/2}$ breaks modular invariance. 
Apart from the implication that every modular invariant theory has a linearly realized $\mathds{Z}_4^R$ symmetry, this means that $\tau$ has to deviate from the critical points once \ac{SUSY} is broken.

Our framework can be naturally extended to theories with several moduli. A detailed discussion will be carried out elsewhere. 
Our results demonstrate that the strong $\CP$ problem is not necessarily an isolated infrared problem: modular consistency ties it to the quark determinant, gauge thresholds, and hidden strong dynamics. 
In particular, the small \ac{QCD} angle and the quark hierarchy can arise as correlated consequences of the same ultraviolet structure.

\begin{acknowledgments}
We thank Mu-Chun Chen for useful discussions. 
X.-G.~L.\ was supported by the Universidad Nacional Aut\'{o}noma de M\'{e}xico Postdoctoral Program (POSDOC). 
This material is based upon work supported by the National Science Foundation Graduate Research Fellowship Program under Grant No.\ DGE-2235784. 
Any opinions, findings, and conclusions or recommendations expressed in this material are those of the authors and do not necessarily reflect the views of the National Science Foundation. 
\end{acknowledgments}

\bibliographystyle{apsrev4-1}
\bibliography{ModularStrongCP}

\clearpage
\begin{center}
{\large\bfseries Supplemental Material}
\end{center}
\setcounter{secnumdepth}{2}
\setcounter{section}{0}
\renewcommand{\thesection}{\Alph{section}}

\section{Modular Forms and the divisor Condition}\label{app:modular-forms}

The modular group is generated by
\begin{subequations}\label{eq:ST-generators}
\begin{align}
\mathsf{S}&=
\begin{pmatrix}0&1\\-1&0\end{pmatrix}~\text{ and }~
\mathsf{T}=
\begin{pmatrix}1&1\\0&1\end{pmatrix}\;,
\label{eq:ST-matrices}
\end{align}
which fulfill
\begin{align}
\mathsf{S}^4&=(\mathsf{S}\mathsf{T})^3=\mathds{1}~\text{ and }~
\mathsf{S}^2 \mathsf{T}=\mathsf{T}\mathsf{S}^2\;.
\label{eq:ST-relations}
\end{align}
\end{subequations}
A $d$-dimensional \ac{VVMF} is a holomorphic vector $Y(\tau)=(Y_1(\tau),\ldots,Y_d(\tau))^{\mathsf T}$ that transforms with modular weight $k$ and a $d$-dimensional representation $\rho$ of $\Group{SL}{2,\mathds{Z}}$,
\begin{equation}\label{eq:VVMF-definition}
Y(\gamma\,\tau)=j_\gamma(\tau)^k\,\rho(\gamma)\,Y(\tau)\;.
\end{equation}
Here $k$ fixes the scalar automorphy factor, whereas $\rho(\gamma)$ mixes the components. 
We restrict to finite-image representations, for which $\rho$ may be chosen unitary. 
In a basis with
\begin{equation}\label{eq:T-diagonal-VVMF}
\rho(\mathsf T)=\diag\left(\ee^{2\piup\ii r_1},\ldots,\ee^{2\piup\ii r_d}\right)\;, \quad r_i\in\mathds{Q}
\end{equation}
holomorphy at the cusp implies
\begin{equation}\label{eq:VVMF-q-expansion}
Y_j(\tau)=q^{r_j}\sum_{n\geq0}a_j(n)q^n\;,\qquad q=\ee^{2\piup\ii\tau}\;.
\end{equation}
The space of \acp{VVMF} in representation $\rho$ is a free module of rank $d$ over the ring $\mathds{C}[E_4,E_6]$~\cite{marks2010structure}.
For a one-dimensional finite-image representation, the space of \acp{VVMF} is generated by $\eta^2$, $E_4$, and $E_6$~\cite{marks2010structure,Liu:2021gwa}.
That is, a form of weight $k$ can be written as a linear combination of monomials 
\begin{equation}\label{eq:one-dimensional-ring}
\eta^{2c}\,E_4^a\,E_6^b\;,
\end{equation}
where $4a+6b+c=k$ and $a,b,c\in\mathds{N}_0$.
The three generators $\eta^2$, $E_4$, and $E_6$ have their fundamental zeros at $\tau=\ii\infty$, $\tau=\ee^{2\piup\ii/3}$, and $\tau=\ii$, respectively. 
The valence formula therefore implies that a one-dimensional form of weight $k$ whose only zero is at the cusp is~\cite{Bruinier:2008xxx,marks2010structure}
\begin{equation}\label{eq:maximal-cusp-form}
  Y(\tau)\propto\eta(\tau)^{2k}\;.
\end{equation}
For $k=0$ this reduces to a constant. 
On the other hand, for $k>0$ it has cusp order $k/12$. 
This is the mathematics underlying \Cref{eq:cusp-form-condition}. 
If the threshold or $\mathscr{W}_0$ has an elliptic-point divisor, additional factors of $E_4$ or $E_6$ are possible, but the full product in \Cref{eq:divisor-condition} must still have zero total divisor.

For general high-dimensional representations, the generators of the space of \acp{VVMF} are described systematically by modular linear differential equations, whose solutions can be expressed through hypergeometric functions for dimensions two and three~\cite{Liu:2021gwa}. 
However, in dimensions four and higher, there is no comparably universal closed formula for an arbitrary representation. 
Instead, they are commonly constructed case by case from combinations of theta constants and eta functions. 
The $\Delta(384)$ sextets used in the main text provide an explicit higher-dimensional example: $Y^{(2)}_{\rep 6_0}$ and $Y^{(2)}_{\rep 6_5}$ are quartic polynomials in level-16 theta constants. 
Their explicit expressions and Fourier expansions are given in~\cite{CentellesChulia:2026bkr}. 
Substituting them into \Cref{eq:model-Yukawas} yields the identity \eqref{eq:model-determinant}.

\section{Supergravity Bookkeeping}\label{app:sugra-bookkeeping}

The main text uses the compensating coefficient $\mathcal{A}_a$ required in the gauge kinetic function. 
For $\Group{SU}{3}_\mathrm{C}$ and the \ac{MSSM} quark multiplets, \Cref{eq:anomaly-coefficient} reduces to
\begin{equation}\label{eq:A3-supplement}
  \mathcal{A}_3=
  \sum_{i=1}^{3}\bigl(2\,k_{Q_i}+k_{U_i^c}+k_{D_i^c}\bigr)
  -C_3\,k_{\mathscr{W}}\;,
\end{equation}
where $C_3=3$.
The term $-C_3\,k_{\mathscr{W}}$ is the net K\"ahler/super-Weyl contribution after the $\Group{SU}{3}_\mathrm{C}$-charged matter and gluino rotations are combined. 

Combining \Cref{eq:A3-supplement,eq:det-weight} with the Higgs condition gives, in the cusp-only class,
\begin{align}
\kappa_3\,\delta_\mathrm{GS}&=\mathcal{K}_Y-\mathcal{A}_3 \notag\\
 &=3(k_{H_u}+k_{H_d})-3k_{\mathscr{W}}=-3k_{\mathscr{W}}\;.\label{eq:supp-delta-relation}
\end{align}
Given that $k_{\mathscr{W}}=1$, our explicit model has $\delta_\mathrm{GS}=-3$ and $C_h=3$, see the discussion around \Cref{eq:model-f3}.

The Yukawa determinant transform as
\begin{equation}\label{eq:det-transform-supplement}
 \det\bigl(Y^u\,Y^d\bigr)(\gamma\,\tau)
 =j_\gamma(\tau)^{\mathcal{K}_Y}\,
  \chi_Y(\gamma)\,
  \det\bigl(Y^u\,Y^d\bigr)(\tau)\;,
\end{equation}
with $\chi_Y=(\chi_3^\rho)^{-1}
 (\chi_{H_u}\chi_{H_d})^{-3}$.
Using \Cref{eq:instanton-transform,eq:A3-supplement,eq:det-weight}, one obtains
\begin{align}
\MoveEqLeft\ee^{-8\piup^2 f_3(S,\gamma\,\tau)}
\det\bigl(Y^uY^d\bigr)(\gamma\,\tau)\notag\\
&=j_\gamma(\tau)^{3(k_{H_u}+k_{H_d})-C_3k_{\mathscr{W}}}
(\chi_{H_u}\chi_{H_d})^{-3}\notag\\
&\quad{}\cdot
\ee^{-8\piup^2 f_3(S,\tau)}
\det\bigl(Y^u\,Y^d\bigr)(\tau)\;.
\end{align}
The vacuum superpotential carries weight $-k_{\mathscr{W}}$, so $\mathscr{W}_0^{-C_3}$ cancels the remaining K\"ahler-$\Group{U}{1}_\mathrm{R}$ weight. 
This gives \Cref{eq:A-transform} and explains why the rigid quantity $\ee^{-8\piup^2 f_3}\,\det(Y^uY^d)$ is incomplete in local \ac{SUGRA}. 
The phase of the resulting invariant $\ASUGRA$ equals $\bar\theta$ once the gluino mass is phase aligned with $\overline{\mathscr{W}}_0$. 
We will discuss this condition in the next section.

Note that canonical normalization cannot generate a new phase. 
If the matter metric is written as $K_{I\bar J}=(Z^\dagger Z)_{I\bar J}$, the field redefinition to canonical variables multiplies the determinant in \Cref{eq:A-SUGRA} by a positive real function. 
This proves \Cref{eq:kahler-correction}.

\section{Gluino Phase Alignment and the hidden Condensate}\label{app:phase-hidden}

The remaining ingredient needed to identify $\arg\ASUGRA$ with the physical $\bar\theta$ is the gluino phase. 
In Einstein frame, the tree-level gaugino mass is
\begin{subequations}\label{eq:supp-gaugino-mass}
\begin{align}
M_a^\mathrm{tree}&=\frac{1}{2\re f_a}F^I\partial_If_a\;,\\
F^I&=-\ee^{\mathscr{K}/2}K^{I\bar J}\,D_{\bar J}\overline{\mathscr{W}}\;.  
\end{align}
\end{subequations}
Let us write $\mathscr{W}_0=\lvert\mathscr{W}_0\rvert\,\ee^{\ii\phi_W}$. 
If, in a $\CP$ basis, $K^{I\bar J}$, $D_I\mathscr{W}/\mathscr{W}$, and $\partial_If_a$ are real along the supersymmetry-breaking directions, then
\begin{subequations}\label{eq:supp-Fphase}
\begin{align}
 D_{\bar J}\overline{\mathscr{W}}&=d_{\bar J}\,\ee^{-\ii\phi_W}\;,\\
 F^I&=r^I\overline{\mathscr{W}}_0\quad\text{where } d_{\bar J},r^I\in\mathds{R}\;.
\end{align}
\end{subequations}
Consequently, $M_a^\mathrm{tree}=r_a\,\overline{\mathscr{W}}_0$ with real $r_a$, and $\arg M_a=-\arg\mathscr{W}_0$ modulo $\piup$, provided all threshold contributions carry the same phase.
Multiple misaligned auxiliary fields or a complex uplift sector contribution invalidate this relation and reintroduce the usual supersymmetric $\CP$ problem.

For a pure $\mathcal{N}=1$ super-Yang--Mills group $G_h$ with dual Coxeter number $C_h$, the holomorphic beta-function coefficient is $b_0=3C_h$. 
The renormalization-group invariant scale obeys 
\begin{equation}\label{eq:supp-RGscale}
\Lambda_h^{3C_h}=M^{3C_h}\exp[-8\piup^2f_h(M)]\;.
\end{equation}
This leads to the non-perturbative superpotential term
\begin{equation}\label{eq:supp-condensate}
\mathscr{W}_\mathrm{np}=A_h\,\exp\left(-\frac{8\piup^2}{C_h}f_h\right)\;.
\end{equation}
The complete hidden Wilsonian function, including any threshold, must appear in this expression. 
For
\begin{equation}\label{eq:supp-general-hidden-f}
  f_h=\kappa_h\,S+\frac{\widetilde{B}_h}{8\piup^2}\,\ln\eta^2+f_h^\mathrm{inv}\;,
\end{equation}
one obtains
\begin{equation}\label{eq:supp-general-hidden-W}
 \mathscr{W}_\mathrm{np}=A_h\,\ee^{-8\piup^2\kappa_hS/C_h}\,
 \eta^{-2\widetilde{B}_h/C_h}\,\ee^{-8\piup^2f_h^\mathrm{inv}/C_h}\;.
\end{equation}
The pure SYM anomaly equation $C_hk_{\mathscr{W}}=\widetilde{B}_h-\kappa_h\delta_\mathrm{GS}$ ensures the required superpotential weight.
Further, the finite multiplier acts consistently on the $C_h$ condensate branches. 
If no additional divisors occur, cancellation of the complete $S$ and cusp dependence implies that
\begin{equation}\label{eq:supp-general-hidden-matching}
\widetilde{B}_3=\mathcal{K}_Y+\frac{C_3}{C_h}\widetilde{B}_h\;,
\quad\text{where }
\kappa_3=\frac{C_3}{C_h}\kappa_h\;.
\end{equation}
This relation ensures that the $\eta$ powers carried by $\ee^{-8\piup^2f_3}$, $\det(Y^u\,Y^d)$, and $\mathscr{W}_0^{-C_3}$ cancel exactly. 
A hidden threshold therefore changes how the cusp divisor is distributed among these factors, but does not generate an uncanceled divisor in $\ASUGRA$.

Taking, for simplicity, $f_h=S$, $\widetilde{B}_h=0$, $\kappa_h=1$, and $C_h=-\delta_\mathrm{GS}=3$ leads to
\begin{equation}\label{eq:supp-W0}
\mathscr{W}_0=A\,\ee^{-8\piup^2S/3}\;.
\end{equation}
As $\delta_\mathrm{GS}=-3$, $S\mapsto S+3(8\piup^2)^{-1}\ln j_\gamma$. 
This means that \Cref{eq:supp-W0} transforms as $j_\gamma^{-1}\,\mathscr{W}_0$ and has trivial finite character, as required for the complete superpotential. 
Within the threshold-free, level-one, single-condensate class, this form is unique up to an overall real constant and the $C_h$ pure-SYM branches. 
Since $C_h=C_3=3$, $\mathscr{W}_0^{-C_3}$ is branch independent. A generic heterotic compactification may instead have $\widetilde{B}_h\neq0$; the minimal choice $\widetilde{B}_h=0$ requires a vanishing hidden $\mathcal{N}=2$ beta-function coefficient or an equivalent cancellation.

\section{Holomorphic Green--Schwarz Frames}\label{app:GS-frame}

In what follows, ``frame'' refers only to a choice of holomorphic chiral-dilaton variable and is unrelated to the Einstein- versus string-frame choice of spacetime metric. 
The main text uses the explicit GS frame in which the continuous Green--Schwarz transformation appears directly as a shift of $S$.

Let
\begin{subequations}\label{eq:supp-eta-log}
\begin{align}
  L_\gamma&=\ln j_\gamma(\tau)\;,\\
  \ell_\eta&=\ln\chi_\eta(\gamma)\;,\\
  \ln\eta^2(\gamma\,\tau)&=\ln\eta^2(\tau)+L_\gamma+\ell_\eta\;.
\end{align}
\end{subequations}
Starting from
\begin{subequations}\label{eq:supp-GS-original}
\begin{align}
  S' &=S-\frac{\delta_\mathrm{GS}}{8\piup^2}L_\gamma \;,\\
  f_a&=\kappa_a\,S+\frac{\widetilde{B}_a}{8\piup^2}\ln\eta^2\;,
\end{align}
\end{subequations}
consider the holomorphic redefinition
\begin{equation}\label{eq:supp-Slambda}
S_\lambda=S+\frac{\lambda}{8\piup^2}\ln\eta^2\;.
\end{equation}
$S_\lambda$ transforms as
\begin{equation}\label{eq:supp-Slambda-transform}
 S_\lambda'=S_\lambda-
 \frac{\delta_\mathrm{GS}-\lambda}{8\piup^2}L_\gamma
 +\frac{\lambda}{8\piup^2}\ell_\eta\;.
\end{equation}
The parameter $\lambda$ therefore labels equivalent holomorphic frames: $\lambda=0$ is the explicit GS frame used in the main text, while nonzero $\lambda$ redistributes the continuous modular shift between the dilaton and the explicit $\eta$ threshold. 
The last term in \Cref{eq:supp-Slambda-transform} is the finite shift induced by the Dedekind-$\eta$ multiplier. 
The gauge kinetic function thus becomes
\begin{equation}\label{eq:supp-frame-data}
f_a=\kappa_a\,S_\lambda+
\frac{B_a^{(\lambda)}}{8\piup^2}\ln\eta^2\;,\quad\text{where }
B_a^{(\lambda)}=\widetilde{B}_a-\kappa_a\lambda\;.
\end{equation}
Its variation under the modular transformation is
\begin{equation}\label{eq:supp-frame-variation}
\Delta_\gamma f_a=
\frac{\mathcal{A}_a}{8\piup^2}L_\gamma+
\frac{\widetilde{B}_a}{8\piup^2}\ell_\eta\;,
\end{equation}
with the frame-independent conditions
\begin{subequations}\label{eq:supp-frame-conditions}
\begin{align}
  &\mathcal{A}_a=B_a^{(\lambda)}-\kappa_a(\delta_\mathrm{GS}-\lambda)\;,\\
  &\chi_a^\rho\chi_\eta^{B_a^{(\lambda)}+\kappa_a\lambda}
  =\chi_a^\rho\chi_\eta^{\widetilde{B}_a}=1\;.\label{eq:supp-frame-condition-b}
\end{align}
\end{subequations}
The finite shift in \Cref{eq:supp-Slambda-transform} is essential. 
Omitting it would incorrectly replace $\widetilde{B}_a$ by $B_a^{(\lambda)}$ in \Cref{eq:supp-frame-condition-b}.

The unique minimal choice that removes the continuous $\ln j_\gamma$ shift is
\begin{subequations}\label{eq:supp-Shat}
\begin{align}
  \widehat{S}&=S+\frac{\delta_\mathrm{GS}}{8\piup^2}\ln\eta^2\;,\\
  \widehat{S}&\longmapsto\widehat{S}+
  \frac{\delta_\mathrm{GS}}{8\piup^2}\ln\chi_\eta\;.  
\end{align}
\end{subequations}
Thus $\widehat{S}$ is not fully invariant. 
Rather, it carries an induced finite shift.
In the $\widehat{S}$ frame
\begin{subequations}\label{eq:supp-hat-frame}
\begin{align}
  f_a&=\kappa_a\widehat{S}+\frac{\mathcal{A}_a}{8\piup^2}\ln\eta^2\;,\\
  1&=\chi_a^\rho\chi_\eta^{\mathcal{A}_a+\kappa_a\delta_\mathrm{GS}}\;.
\end{align}  
\end{subequations}
For the explicit model,
\begin{subequations}\label{eq:supp-model-hat}
\begin{align}
  \widehat{S}&=S-\frac{3}{8\piup^2}\ln\eta^2\;,\\
  f_3&=\widehat{S}+\frac{15}{8\piup^2}\ln\eta^2\;,\\
  \mathscr{W}_0&=A\ee^{-8\piup^2\widehat{S}/3}\eta^{-2}\;.
\end{align}
\end{subequations}
The multiplier of $\eta^{-2}$ is canceled by the finite shift of $\widehat{S}$, so the complete superpotential still has trivial finite character. 
Furthermore,
\begin{subequations}\label{eq:supp-frame-cancellation}
\begin{align}
  \ee^{-8\piup^2f_3}\det(Y^uY^d)
  &\propto\ee^{-8\piup^2\widehat{S}}(\eta^2)^{-3}\;,\\
  \mathscr{W}_0^{-3}
  &=A^{-3}\ee^{8\piup^2\widehat{S}}(\eta^2)^3\;,
\end{align}
\end{subequations}
which reproduces the same constant as \Cref{eq:model-Aconstant}. 
That is, the two frames only redistribute the continuous weight, finite multiplier, and cusp divisor among $S$, $f_3$, and $\mathscr{W}_0$.

\section{UV--IR invariance of \texorpdfstring{$\bar\theta$}{theta-bar}}\label{app:UVIR}

Add a vector-like pair $X$ and $\bar X$ in the fundamental and antifundamental representations of $\Group{SU}{3}_\mathrm{C}$ with
\begin{equation}\label{eq:heavy-mass-term}
  \mathscr{W}_{\mathrm{UV}}\supset M(\tau)\,X\,\bar X\;.
\end{equation}
Modular covariance requires that
\begin{subequations}\label{eq:heavy-mass-weight}
\begin{align}
  M(\gamma\,\tau)&=j_\gamma(\tau)^{k_M}\,\chi_M(\gamma)\,M(\tau)\;,\\
  k_M&=k_X+k_{\bar X}-k_{\mathscr{W}}\;.  
\end{align}
\end{subequations}
For one fundamental pair, the anomaly coefficient satisfies
\begin{equation}\label{eq:anomaly-gap}
\mathcal{A}_3^{\mathrm{UV}}
=\mathcal{A}_3^{\mathrm{light}}+k_M\;.
\end{equation}
The holomorphic threshold is
\begin{equation}\label{eq:IR-threshold-supplement}
f_3^{\mathrm{IR}}(S,\tau)
=f_3^{\mathrm{UV}}(S,\tau)
-\frac{1}{8\piup^2}\ln M(\tau)\;.
\end{equation}
Its automorphy factor changes the coefficient from $\mathcal{A}_3^{\mathrm{UV}}$ to $\mathcal{A}_3^{\mathrm{light}}$, while the multiplier $\chi_M$ removes the heavy finite determinant character. 
Therefore, both \Cref{eq:continuous-layer,eq:character-layer} are satisfied.

The full holomorphic determinant from $\Group{SU}{3}_\mathrm{C}$-charged states above the threshold is
\begin{equation}\label{eq:UV-determinant}
\det\mathcal M_{\mathrm{UV}}
=M(\tau)\,\det\bigl(Y^uY^d\bigr)_{\mathrm{IR}}\;.
\end{equation}
Exponentiating \Cref{eq:IR-threshold-supplement} gives
\begin{equation}\label{eq:exponentiated-threshold}
\ee^{-8\piup^2 f_3^{\mathrm{IR}}}
=M(\tau)\,\ee^{-8\piup^2 f_3^{\mathrm{UV}}}\;.
\end{equation}
Consequently,
\begin{align}
  A_{\mathrm{UV}}
  &=\ee^{-8\piup^2 f_3^{\mathrm{UV}}}\,
  M(\tau)\,\det\bigl(Y^u\,Y^d\bigr)_{\mathrm{IR}}\,
  \mathscr{W}_0^{-C_3}\notag\\
  &=\ee^{-8\piup^2 f_3^{\mathrm{IR}}}\,
  \det\bigl(Y^u\,Y^d\bigr)_{\mathrm{IR}}\,
  \mathscr{W}_0^{-C_3}
  =A_{\mathrm{IR}}\;.\label{eq:A-UVIR}
\end{align}
Hence,
\begin{equation}
\label{eq:theta-UVIR}
\bar\theta_{\mathrm{UV}}=\bar\theta_{\mathrm{IR}}\;.
\end{equation}
Integrating out a modular-covariant vector-like pair merely moves its phase from the mass determinant into $f_3$. 
The equality fails only if the heavy mass contains an independent $\CP$ phase which is not part of the modular covariant holomorphic quantities.

\end{document}